\documentclass{article}
\usepackage[utf8]{inputenc}

\usepackage{amsmath}

\usepackage{amssymb}

\newtheorem{theorem}{Theorem}
\newtheorem{corollary}[theorem]{Corollary}

\newtheorem{lemma}[theorem]{Lemma}

\newtheorem{definition}[theorem]{Definition}
\newtheorem{proposition}[theorem]{Proposition}
\newtheorem{observation}[theorem]{Observation}

\newenvironment{proof}{\noindent\bf{Proof.}\rm}{\hfill$\blacksquare$\bigskip}

\newcommand{\items}{\mathcal{M}}

\begin{document}

\title{The Residual Maximin Share}

\author{Uriel Feige\thanks{Weizmann Institute, Israel. {\tt uriel.feige@weizmann.ac.il}} }

\maketitle

\begin{abstract}
    We consider fair allocations of indivisible goods to agents with general monotone valuations. We observe that it is useful to introduce a new share-based fairness notion, the {\em residual maximin share} (RMMS). This share is {\em feasible} and {\em self maximizing}. Its value is at least as large as the MXS for monotone valuations, and at least as large as $\frac{2}{3}$-MMS for additive valuations. Known techniques easily imply the existence of partial allocations that are both RMMS and EFX, and complete allocations that are both RMMS and EFL. This unifies and somewhat improves upon several different results from previous papers.  
\end{abstract}

\section{Introduction}

We consider a well studied setting of allocating a set $\items = \{e_1, \ldots, e_m\}$ of $m$ indivisible items (also referred to as goods) to $n$ agents, $a_1, \ldots, a_n$. 
Each agent $i$ has a valuation function $v_i$, where valuations are assumed to be normalized ($v_i(\emptyset) = 0$) and monotone ($v_i(S) \le v_i(T)$ for every $S \subset T \subseteq \items$). A partial allocation ${\mathcal{A}} = (A_0, A_1, \ldots, A_n)$ is a partition of $\items$ into $n+1$ bundles, such that for every $i \ge 1$, agent $a_i$ receives bundle $A_i$. If $A_0$ is empty, then we refer to the partial allocation as a full allocation, or simply an allocation. We wish allocations to satisfy some fairness criteria. 

There are two major classes of fairness criteria for allocations. One is share-based fairness, in which an agent judges an allocation to be fair only based on the bundle of items that she herself receives. Examples of shares considered in this paper are the MMS (maximin share)~\cite{Budish11}, and the MXS (minimum EFX share)~\cite{CaragiannisGRSV23}. The other class is comparison-based fairness, in which an agent judges whether an allocation is fair by comparing the bundle that she receives with bundles received by other agents. Examples considered in this paper are EFX (envy free up to any item)~\cite{CKMPSW19}, EFL (envy free up to one less preferred good)~\cite{BarmanBMN18}, and EF1 (envy free up to one item)~\cite{Budish11}. See Section~\ref{sec:preliminary} for definitions of these and other concepts mentioned in the introduction.

For a share $s$ (such as MMS, MXS), an $s$-allocation (partial $s$-allocation, respectively) is an allocation (partial allocation, respectively) in which every agent gets a bundle that she values at least as high as her $s$ share value. A share $s$ is {\em feasible} for a class $C$ of valuations if in every allocation instance with valuations from the class $C$, there is an $s$-allocation. For a given ratio $0 < \rho < 1$ and a share $s$,  $\rho$-$s$ is a share whose value is $\rho$ times the value of the $s$ share. 

We introduce a new share, that we refer to as the {\em  residual maximin share} (RMMS). Before presenting it, we explain the notion of residual self-feasibility. This notion was made use of in previous work~\cite{KPW18, amanatidis2017approximation, AkramiR25a}, though the terminology that we use is new.

For a given share $s$ and valuation $v$, a partition $P_1, \ldots, P_n$ of $\items$ will be referred to as a {\em $v$-acceptable $n$-partition} if $v(P_j) \ge s(\items, v, n)$ for every $j$. We say that a share $s$ is {\em self-feasible} for class $C$ (and a given value of $n$) if every $v \in C$ has a $v$-acceptable $n$-partition.  A share is {\em residual self-feasible} if for every $v \in C$, every $0 \le k < n$, and every $k$ bundles each of value strictly less than $s(\items, v_i, n)$, after removing these bundles there is a $v$-acceptable $n-k$ partition (of the set of remaining items). 

The MMS is the self-feasible share with highest possible value. We introduce the RMMS share, which is the residual self-feasible share with highest possible value.

    \begin{definition} 
    \label{def:RSF}
    The value of the {\em residual maximin share} (RMMS) for valuation $v$, denoted as $RMMS(\items, v, n)$, is the highest value $t$ that is residual self-feasible. That is, for every $0 \le k < n$, removing $k$ bundles each of value (under $v$) strictly less than $t$, there is an $(n-k)$ partition of the set of remaining items, where each part has value (under $v$) at least $t$. 
\end{definition}

By definition, the RMMS is never larger than the MMS. By design, the RMMS is at least as large as the MXS (see Definition~\ref{def:MXS}).

\begin{observation}
    \label{obs:MXS}
    For every monotone valuation, the RMMS is at least as large as the MXS.
\end{observation}

(All proofs appear in section~\ref{sec:proofs}.)

Being residual self-feasible implies that a share is feasible. This was demonstrated in several special cases. see for example~\cite{amanatidis2017approximation, AkramiR25}.

\begin{observation}
\label{obs:RSFfeasible}
    The RMMS is {\em feasible} for the class of monotone valuations. 
\end{observation}

The proof of Observation~\ref{obs:RSFfeasible} will be omitted, because in Observation~\ref{obs:EFXRSF} we shall prove an even stronger statement.

Hence like MXS, the RMMS is feasible for all monotone valuations. It enjoys the advantage of being at least as large as the MXS. Here we establish another aspect in which RMMS is preferable over MXS. MXS is a somewhat unnatural share, as it is not monotone -- increasing the values of some bundles might decrease the value of the share. In contrast, RMMS is monotone, and moreover, is {\em self-maximizing}~\cite{BF22} (see Definition~\ref{def:selfMaximizing}), a property that implies several other natural properties (such as monotonicity, and being 1-Lipschitz).  More details appear in Section~\ref{sec:preliminary}.

\begin{proposition}
\label{pro:RSFselfMaximizing}
    The RMMS is a {\em self-maximizing} share. 
\end{proposition}

For some natural restricted classes of valuations, the RMMS offers nontrivial guarantees with respect to the MMS.

\begin{observation}
\label{obs:RSFdominating}
    For unit-demand valuations, the RMMS equals the MMS, for additive valuations it is at least $\frac{2}{3}$-MMS (the ratio $\frac{2}{3}$ can be replaced by $\frac{2n}{3n-1}$ for odd $n$ and $\frac{2n-2}{3n-4}$ for even $n$), and for subadditive valuations, it is at least $\frac{1}{n}$-MMS. 
\end{observation}

In comparison, for additive valuations the value of MXS is at least $\frac{4}{7}$-MMS, but sometimes smaller than $\frac{3}{5}$-MMS~\cite{ABM18}. 

In passing, we note that the combination of Observations~\ref{obs:RSFfeasible} and~\ref{obs:RSFdominating} 
implies the following corollary.

\begin{corollary}
    \label{cor:subadditive}
    For agents with subadditive valuations,  $\frac{1}{n}$-MMS allocations always exist.
\end{corollary} 

Our proof for Observation~\ref{obs:RSFdominating} actually shows that every EF1 allocation gives agents with subadditive valuations at least $\frac{1}{n}$-MMS. Hence, Corollary~\ref{cor:subadditive} is also a consequence of the existence of EF1 allocations~\cite{LiptonMMS04}, or the existence of MXS allocations~\cite{AkramiR25}. We are not aware of the corollary being stated in previous work.
The bound stated in Corollary~\ref{cor:subadditive} holds for all $n$. For sufficiently large  $n$, and also for small $n$, better bounds are known. See Section~\ref{sec:related} for more details.

We further observe that a proof technique of~\cite{AkramiR25a} implies feasibility of RMMS simultaneously with satisfying a strong comparison-based fairness notion.

\begin{observation}
\label{obs:EFXRSF}
    In every allocation instance with monotone valuations, there is a partial RMMS-allocation that is EFX.
\end{observation}

As a corollary, we obtain the following.

\begin{corollary}
    \label{cor:EFLRSF}
    In every allocation instance with monotone valuations, there is a full RMMS-allocation that is EFL.
\end{corollary}

The proof of this corollary is by a straightforward extension of known techniques. (Had we replaced EFL with the weaker EF1 property, a note in the discussion section of~\cite{CaragiannisGH19} would imply the corollary. The fact that EF1 can be replaced by EFL is implicit in some previous work.)

Previously, it was shown that for additive valuations, there are partial allocations that are EFX and $\frac{2}{3}$-MMS and full allocations that are EFL and $\frac{2}{3}$-MMS~\cite{AkramiR25a} (the result there is stated for EF1, not EFL, but the proofs do give EFL), and that for {\em restricted MMS-feasible} valuations (a class of valuations that is more general than additive valuations) simultaneous EFL and MXS allocations exist~\cite{AG24}. The combination of Corollary~\ref{cor:EFLRSF} and observations~\ref{obs:MXS}, \ref{obs:RSFdominating} and~\ref{obs:EFXRSF} unify and somewhat extend these results. This is summarized in the following corollary.

\begin{corollary}
    \label{cor:unify}
    In every allocation instance with monotone valuations, there is a partial MXS-allocation that is EFX, and a full MXS-allocation that is EFL. Moreover, 
    if valuations are additive, these guarantees are satisfied together with $\frac{2}{3}$-MMS.  
\end{corollary}

\section{Preliminaries}
\label{sec:preliminary}

We assume that $m > n$, as otherwise we allocate all items, giving each agent at most one item, and this allocation will satisfy all fairness notions considered in this paper.

For a set $S \subseteq \items$ and an item $e \in S$, we use $S - e$ as shorthand notation for $S \setminus \{e\}$.

A valuation function $v$ is {\em additive} if for every set $S \subset \items$, $v(S) = \sum_{e \in S} v(\{e\})$. It is {\em subadditive} if for every $S, T \subset \items$, $v(S) + v(T) \ge v(S \cup T)$. It is {\em unit demand} if for every set $S \subset \items$, $v(S) = \max_{e \in S} v(\{e\})$. 

We review some comparison-based fairness notions. Given a partial allocation $A_0, A_1, \ldots, A_n$, we consider four types of ``envy" of agent $a_i$ towards agent $a_j$. We list them from weakest envy (with the complement property being the strongest form of {\em envy-freeness}) to strongest envy (weakest envy-freeness).

\begin{enumerate}
\item EF envy: $v_i(A_i) < v_i(A_j)$.
    \item EFX envy: for some item $e \in A_j$, $v_i(A_i) < v_i(A_j -e)$.
    \item EFL envy: $|A_j| \ge 2$, and for every item $e \in A_j$, either $v_i(A_i) < v_i(e)$ or $v_i(A_i) < v_i(A_j -e)$.
        \item EF1 envy: for every item $e \in A_j$, $v_i(A_i) < v_i(A_j -e)$.
\end{enumerate}

\begin{definition}
    A partial allocation is EF (EFX, EFL, EF1, respectively) if there is no pair of agents $a_i$ and $a_j$ for which $a_i$ experiences EF envy (EFX envy, EFL envy, EF1 envy, respectively) towards $a_j$.
\end{definition}

It follows from the definitions that every EFX allocation is EFL, and every EFL allocation is EF1.

We now review some share-based fairness notions.

\begin{definition}
    Given an allocation instance with $n$ agents, the MMS share of agent $a_i$, denoted by $MMS(\items, v_i, n)$, is the maximum over all partitions $P_1, \ldots, P_n$ of $\items$ of the minimum value of a bundle in the partition, $\min_{j} v_i(P_j)$. 
\end{definition}

\begin{definition}
\label{def:MXS}    Given an allocation instance with $n$ agents, the MXS share of agent $a_i$, denoted by $MXS(\items, v_i, n)$, is the minimum value $v_i(S)$ over all bundles $S \subset \items$ for which there is an allocation $A_1, \ldots, A_n$ with $A_i = S$ in which agent $a_i$ has no EFX envy towards any other agent.
\end{definition}

It follows from the definitions that every MMS allocation is an MXS allocation.

A systematic study of shares was initiated in~\cite{BF22}. There (and also in some earlier work), several properties that are desirable for shares were presented. One of them was that of being feasible, and we list here two other natural properties.

\begin{definition}
    \label{def:natural}
    We say that valuation $v_i$ dominates valuation $v_j$ if for every bundle $S$, $v_i(S) \ge v_j(S)$. For a given $\epsilon > 0$, we say that two valuations are $\epsilon$-close if for every bundle $S$, $|v_i(S) - v_j(S)| \le \epsilon$. A share is {\em monotone} if whenever $v_i$ dominates $v_j$, the share value for $v_i$ is at least as high as that for $v_j$. A share is {\em 1-Lipschitz} if whenever $v_i$ and $v_j$ are $\epsilon$-close, their share value differs by at most $\epsilon$.
\end{definition}

As an example, MMS is both monotone and 1-Lipschitz, whereas $\frac{2}{3}$-MMS is monotone but not 1-Lipschitz (and not even continuous, see~\cite{BF22}), and MXS is not monotone. 

Being {\em self-maximizing} is a property of shares that implies both monotonicity and 1-Lipschitz (see~\cite{BF22}), and also relates to incentives for agents to report their true valuations. (If an agent is risk averse and wishes to maximizes the minimum possible value that she might get by an allocation mechanism that only ensures that she gets at least her share value, she has no incentive to report a valuation $v'$ instead of her true valuation $v$.)

\begin{definition}
    \label{def:selfMaximizing}
    We say that a share $s$ is self maximizing if for every two valuations $v$ and $v'$, there is a bundle $T$ satisfying $v'(T) \ge s(\items, v', n)$ (hence, $T$ is feasible for $v'$ under share $s$) such that $v(T) \le s(\items, v, n)$ (hence, even the worst bundle that is feasible under the share $v$ is at least as valuable as $T$, according to the true valuation $v$ of the agent).
\end{definition}

The MMS is a self-maximizing share, but is not feasible, not even for additive valuations~\cite{KPW18}.  For the special case of additive valuations, the {\em nested share} of~\cite{BF22} is both feasible and self maximizing, and its value is at least $\frac{2n}{3n-1}$-MMS. (Moreover, it can be computed in polynomial time, and so can a feasible allocation.)

Moving to comparison-based fairness notions, 
Lipton, Markakis, Mossel and Saberi~\cite{LiptonMMS04} designed an allocation algorithm that produces EF1 allocations. We refer to their algorithm as the LMMS algorithm. The LMMS algorithm starts from the empty allocation, and then allocates items one by one. However, as observed in multiple previous works, it can also be started from any partial allocation. The following lemma (whose proof is omitted, as it follows trivially from~\cite{LiptonMMS04})  summarizes its properties in this case.

\begin{lemma}
    \label{lem:LMMS}
    Let ${\mathcal P} = (P_0, P_1, \ldots, P_n)$ be a partial allocation, and let ${\mathcal A} = (A_1, \ldots, A_n)$ be the final allocation if the LMMS algorithm is run starting at $P$. Then ${\mathcal A}$ satisfies the following properties:
    \begin{enumerate}
        \item There is a matching $\pi$ between the bundles $A_1, \ldots, A_n$ and $P_1, \ldots, P_n$ such that $P_{\pi(i)} \subset A_i$ for every $i$.
        \item For every agent $a_i$, $v_i(A_i) \ge v_i(P_i)$.
        \item For any $j$, if $A_j \not= P_{\pi(j)}$, let $e_j$ be the last item inserted into $A_j$ by the LMMS algorithm. Then for every agent $i$, $v_i(A_i) \ge v_i(A_j - e_j)$.
    \end{enumerate}
\end{lemma}

\subsection{Additional related work}
\label{sec:related}

To put the results of this paper in context, we present here some additional related work.

As noted in Observation~\ref{obs:RSFdominating}, for additive valuations RMMS is at least $\frac{2}{3}$-MMS. It is known that $\rho$-MMs is feasible for additive valuations for $\rho = \frac{3}{4} + \frac{3}{3836}$~\cite{AG24}, but not feasible for $\rho > \frac{39}{40}$~\cite{FST21}. 

Corollary~\ref{cor:subadditive} shows that $\frac{1}{n}$-MMS is feasible for subadditive valuations. It is known that for subadditive valuations, $\rho$-MMS is not feasible for $\rho > \frac{1}{2}$~\cite{GhodsiHSSY22}, $\frac{1}{2}$-MMS is feasible for $n \le 4$~\cite{cCMS25}, and $\frac{1}{14 \log n}$-MMS is feasible for all $n$~\cite{FH25}.
(Recent work improves the ratio to $\frac{1}{8 \log \log n}$~\cite{SS25,feige25}.) 

Observation~\ref{obs:EFXRSF} constructs partial allocations that are both RMMS and EFX, and Corollary~\ref{cor:EFLRSF} constructs full allocations that are both RMMS and EFL. In this context we note that for subadditive valuations, it is known that there are partial allocations that are EFX and preserve at least half of the Nash Social Welfare~\cite{BarmanS24}, and that they can be extended to full allocations that are EFL and preserve at least half of the Nash Social Welfare. (The last result is stated in~\cite{BarmanS24} with EF1 rather than EFL, but their allocation satisfies the stronger EFL property.)

\section{Proofs}
\label{sec:proofs}

All proofs in this paper are relatively straightforward, given the related work. Hence, our presentation of the proofs will not be very detailed.

\subsection{RMMS dominates MXS: proof of Observation~\ref{obs:MXS}}

\begin{proof}
    The proof of Lemma 3.1 in~\cite{AkramiR25} shows that removing a single bundle whose value is smaller than the MXS and reducing the number of agents by one, the value of the MXS share cannot decrease.  This in combination with the fact that the MXS is self-feasible (its value is never larger than the MMS, which is self-feasible) implies that the MXS is residual self-feasible. Among all shares that satisfy the residual self-feasibility property, RMMS has the highest value. 
\end{proof}

\subsection{RMMS is self maximizing: proof of Proposition~\ref{pro:RSFselfMaximizing}}

\begin{proof}
Consider an arbitrary partition of the set $2^{\items}$ of all $2^m$ possible bundles into two subset, $Y$ and $N$. We say that $Y$ is self-feasible if there is an $n$-partition of $\items$ in which all $n$ bundles are in $Y$. We say that $Y$ is residual self-feasible if for ever $0 \le k < n$ and every $k$ bundles in $N$, removing their items from $\items$, the remaining set of items has an $(n-k)$-partition in which all $n-k$ bundles are in $Y$. 

Given a valuation function $v$, the RMMS value partitions the set $2^{\items}$ of all $2^m$ possible bundles into the subset $Y_v$ of those bundles that are acceptable (have $v$ value at least as high as the share value), and the subset $N_v$ of those bundles that are not acceptable. $Y_v$ is required to be residual self-feasible, and among all possible residual self-feasible $Y$, RMMS dictates that we choose the one in which the lowest value bundle has highest possible value (under $v$). Hence no matter what $v'$ an agent with valuation $v$ reports, the corresponding set of acceptable bundles under $v'$ will necessarily contain at least one bundle $T$ of value $v(T) \le RMMS(\items, v, n)$.
\end{proof}

\subsection{RMMS versus MMS: proof of Observation~\ref{obs:RSFdominating}}
\label{sec:subadditive}

\begin{proof}
For unit-demand valuations, it is not difficult to see that both the MMS and the RMMS are equal to the value of the $n$th most valuable item.

The fact that for additive valuations, $\frac{2}{3}$-MMS  (more precisely, $\frac{2n}{3n-1}$ for odd $n$ and $\frac{2n-2}{3n-4}$ for even $n$) has the residual self-maximizing property was proved in~\cite{KPW18}.

For subadditive valuations, we prove that every EF1 allocation gives each agent at least $\frac{1}{n}$-MMS. It follows that MXS implies $\frac{1}{n}$-MMS, and consequently that also RMMS implies $\frac{1}{n}$-MMS.

Our proof is basically the same as the proof given for the additive case~\cite{CKMPSW19}. 
    
    Consider an arbitrary EF1 allocation $\mathcal{A} = (A_1, \ldots, A_n)$, and suppose that agent $a_n$ has a subadditive valuation $v_n$. Then for every $1 \le j \le n-1$, if $A_j$ is not empty then there is an item $e_j \in A_j$ such that $v_n(A_n) \ge v_n(A_j - e_j)$. Consider now any MMS partition ${\mathcal P} = P_1, \ldots, P_n$. At least one of the parts does not contain any of the $e_j$s. Its value is then at most $v_n(\items \setminus \{e_1, \ldots, e_{n-1}\}) \le A_n + \sum_j v_n(A_j - e_j) \le n\cdot v_n(A_n)$, where the first inequality uses subadditivity of $v_n$.
\end{proof}

\subsection{RMMS and EFX: proof of Observation~\ref{obs:EFXRSF}}

The proof is similar to the proof of Akrami and Rathi~\cite{AkramiR25a} that for additive valuations, there always are allocations that are $\frac{2}{3}$-MMS and EFX. For completeness, we present the proof.

\begin{proof}
We say that a bundle $S$ is desirable for agent $a_i$ if $v_i(S) \ge RMMS(\items, v_i, n)$.

Observe that the RMMS value of an agent might be~0, if fewer then $n$ items have positive value for the agent. This somewhat complicates our presentation, adding distinctions between empty and non-empty bundles.

We present an allocation algorithm that proceeds in rounds.
At the beginning of every round, we have {\em assigned} items (those currently assigned to agents) and {\em free} items (the remaining items). Every agent is either {\em wealthy} (already holds a desired bundle) or {\em poor} (holds no item at all). The algorithm ends when all agents are wealthy. (Proposition~\ref{pro:AR24} will show that indeed such a stage is reached.)

Initially, all items are free. We now describe a single round $r$.

\begin{enumerate}
    \item Let $n_r \ge 1$ denote the number of poor agents, and let $a_i$ be a poor agent. Let $F_r$ denote the set of free items. Construct an arbitrary $v_i$-acceptable $n_r$-partition of $F_r$. This is a collection ${\mathcal P} = (P_1, \ldots, P_{n_r})$ of $n_r$ disjoint bundles such that every $P_j$ is desirable for $a_i$.  (Proposition~\ref{pro:AR24} will show that such a partition must exist.)
    \item For each non-empty bundle $P_j$ in $\mathcal P$, let $P'_j \subseteq P_j$ be a minimal non-empty subset  of $P_j$ (minimal in the sense that no strict subset of $P'_j$ qualifies) so that at least one poor agent desires $P'_j$. Such a $P'_J$ necessarily exists, because $a_i$ desires $P_j$. If there are several possible candidates for $P'_j$, choose one of them arbitrarily. 
    \item If for some bundle $P'_j$ there is a strict subset $S \subset P'_j$ that some wealthy agent values strictly more than her current bundle, then let $S$ be a minimal such subset. The items currently assigned to the respective wealthy agent become free, and instead $S$ is assigned to that agent. This ends the round. 
    \item Else, construct a bipartite graph with poor agents on one side, bundles $P'_1, \ldots, P'_{n_r}$ on the other side, and edges between agents and bundles that they desire. 
    \begin{enumerate}
        \item If this graph has a perfect matching, allocate the bundles to the poor agents according to any such matching, and the algorithm ends.
        \item Else, let $t_r$ be the smallest integer such that there is a set ${\mathcal T}$ of $t_r$ bundles such that there are only $t_r - 1$ poor agents that desire a bundle from ${\mathcal T}$. By Hall's condition, $t_r \le n$, and by the fact that every bundle is desired by at least one poor agent, $t_r \ge 2$. Match each of these $t_r - 1$ poor agents to some bundle in ${\mathcal T}$ (such a matching, leaving one bundle unmatched, must exist, by Hall's theorem and the minimality of $t_r$). Assign items to agents according to this matching, and end the round.
    \end{enumerate}
\end{enumerate}

We now prove correctness of the algorithm.

\begin{proposition}
\label{pro:AR24}
    The algorithm described above terminates.
\end{proposition}

\begin{proof}
    Observe that at every round $r$, for every agent $a_i$ that is poor at the beginning of the round it holds that no bundle that was allocated in a previous round is desirable. The residual self-feasibility property of RMMS then implies that an RMMS partition of $F_r$ exists, as required in step~1 of the algorithm. 

    In every round, either a wealthy agent replaces her bundle by one of higher value (step~3), or at least one poor agent becomes wealthy (step~4). As there are only finitely many different bundles that can be given to an agent (at most $2^m$), step~3 can be executed at most finitely many times (less than $n \cdot 2^m$). Step~4 can be executed at most $n$ times before all agents are wealthy, and then the algorithm ends.
\end{proof}

\begin{proposition}
    The partial allocation is RMMS and EFX.
\end{proposition}

\begin{proof}
    When the algorithm ends, all agents are wealthy, implying that the partial allocation is RMMS.

    Every bundle containing at least two items that is allocated by the algorithm is minimal, in the sense that removing any item from it, no poor agent desires it, and no wealthy agent envies it. 
    (The reason for excluding in the above statement bundles with only one item is because there might be agents whose RMMS value is~0. If so, bundles containing a single item might not be considered to be minimal.) As every agent gets a desirable bundle, removing any item from the bundle of any other agent, the remaining bundle is not envied. Hence, the partial allocation is EFX.
\end{proof}

This completes the proof of Observation~\ref{obs:EFXRSF}.
\end{proof}

\subsection{RMMS and EFL: proof of Corollary~\ref{cor:EFLRSF}}

The corollary follows immediately from applying the following Lemma~\ref{lem:complete} to the partial allocation of Observation~\ref{obs:EFXRSF} (and using the fact that EFX implies EFL).

\begin{lemma}
    \label{lem:complete}
    Let ${\mathcal P} = (P_0, P_1, \ldots, P_n)$ be a partial allocation that is EFL. Then there is a full allocation ${\mathcal A} = (A_1, \ldots, A_n)$ that is EFL, and satisfies $v_i(A_i) \ge v_i(P_i)$ for every agent $a_i$. 
\end{lemma}

The known approach for proving statements similar to Lemma~\ref{lem:complete} (see the discussion section in~\cite{CaragiannisGH19}) is to use the LMMS algorithm and apply Lemma~\ref{lem:LMMS}. Trivially, this implies a variation on Lemma~\ref{lem:complete}, in which the final allocation is EF1 (but not necessarily EFL). Moreover, if in the initial EFL partial allocation no agent envies $P_0$ (as is the case for example in the partial EFX allocation of~\cite{ChaudhuryKMS21}), then the final allocation will be EFL.

Our proof of Lemma~\ref{lem:LMMS} is designed in a way that will later allow us to prove also Proposition~\ref{pro:complete} (see Section~\ref{sec:discuss}). We add a pre-processing step, in which the partial allocation ${\mathcal P} = (P_0, P_1, \ldots, P_n)$ of Lemma~\ref{lem:complete} is replaced by a new partial allocation ${\mathcal P'} = (P'_0, P'_1, \ldots, P'_n)$. The partial allocation ${\mathcal P'}$ will have the property that  there is no free item (item in $P'_0$) that some agent strictly prefers over her own bundle. 

We now present the proof of Lemma~\ref{lem:complete}. 

\begin{proof}
We perform a pre-processing step that proceeds in rounds. In every round, if there is a free item that some agent strictly prefers over her own bundle, one such agent (chosen arbitrarily) takes that item, and gives back her other items (making them free). When every agent weakly prefers her current bundle over every single  free item, the pre-processing step ends. 
The number of rounds is less than $n\cdot m$, even in the most naive implementation of the pre-processing step.

Observe that after the pre-processing step, the resulting partial allocation ${\mathcal P'} = (P'_0, P'_1, \ldots, P'_n)$ is still EFL and satisfies $v_i(P'_i) \ge v_i(P_i)$ for every agent $a_i$.
We use ${\mathcal P'}$ instead of ${\mathcal P}$ as the starting point for the LMMS algorithm.

Lemma~\ref{lem:LMMS} implies that after running the LMMS algorithm, the final allocation ${\mathcal A} = (A_1, \ldots, A_n)$ has the EFL property. Indeed, consider any two agents, $a_i$ and $a_j$. If $A_j$ is a bundle that some agent held in ${\mathcal P'}$, 
then by not having EFL envy in ${\mathcal P'}$, $a_i$ has no EFL envy towards $a_j$ in the ${\mathcal A}$. Else, if $A_j$ contains items that were added by the LMMS algorithm, let $e_j$ be the last item to be added to $A_j$. Then $v_i(A_i) \ge v_i(e)$ (because of the pre-processing step), and $v_i(A_i) \ge v_i(A_j - e)$ (by Lemma~\ref{lem:LMMS}). Hence, the EFL property holds.
\end{proof}

\section{Discussion}
\label{sec:discuss}

The RMMS satisfies three desirable properties: it is feasible (for the class of all monotone valuations), it is self-maximizing, and it has reasonably high value, in particular, at least $\frac{2}{3}$-MMS for additive valuations. We are not aware of any other share that is known to enjoy such a combination of properties. 

This is not to say that the RMMS has no weaknesses. The next subsections present properties of RMMS that can be regarded as weaknesses.

\subsection{RMMS beyond additive valuations}

Recall that the maximin share (MMS) is the value of the worst bundle in the best $n$-partition. One may consider a related share, the {\em minimax share}, defined as the value of the best bundle in the worst $n$-partition. For additive valuations, the minimax share is at least as high as the maximin share, but for other classes of valuations this need not hold. In particular, for submodular valuations (satisfying $v(S + e) - v(S) \ge v(T + e) - v(T)$ for every item $e$ and sets $S \subset T \subset \items$), the value of the minimax share might be as low as $\frac{1}{n}$ times the MMS. This happens when there are $n^2$ items that belong to $n$ groups, each containing $n$ items, and the value of a set $S$ is the number of groups that it intersects. The MMS is $n$ (partition $\items$ into $n$ bundles, each containing one item from each group), whereas the value of the minimax share is only~1 (partition $\items$ into $n$ bundles, each composed of a single group).


\begin{proposition}
\label{pro:minimax}
    For every class of valuations, the value of the RMMS is not larger than that of the minimax share.
\end{proposition}

\begin{proof}
    Consider an arbitrary $n$-partition of $\items$ (or specifically, the one that determines the value of the minimax share). The RMMS value cannot be strictly higher than that of the most valuable bundle in the partition. Otherwise, for RMMS, we would be allowed to remove any $n-1$ bundles of that partition (as their value would be strictly smaller than the RMMS), and the remaining bundle would have to have value at least the RMMS.  
\end{proof}

Proposition~\ref{pro:minimax} implies
that beyond additive valuations, the RMMS does not approximate the MMS very well. 
In particular, for submodular valuations RMMS only ensures $\frac{1}{n}$-MMS, whereas $\frac{10}{27}$-MMS allocations are known to exist~\cite{BUF23}. 



\subsection{Computational aspects}

As is standard when considering arbitrary valuation functions, we assume query access to the valuations. A common query model is that of value queries, in which a query specifies a bundle $S$, and the agent $a_i$ replies with its value $v_i(S)$. An even simpler kind of query model is that of comparison queries, a query model introduced in~\cite{bu2024fair}. In a comparison query, two bundles $S$ and $T$ are presented to an agent $a_i$, and the agent only needs to reply with a single bit, indicating whether $v_i(S) \ge v_i(T)$ or not. Observe that two value queries suffice in order to implement a comparison query, but comparison queries cannot implement value queries.

We assume that values of all bundles are integers, in the range $[0, K]$, implying in particular that $v_i(\items) \le K$. We refer to an algorithm (working under some query model to the valuations, where replying to a query is assumed to take one unit of time) as strongly polynomial time if it runs in time polynomial in $n$, $m$ (independent of $K$), as polynomial time if it runs in time polynomial in $n$, $m$ and $\log K$, and as pseudo polynomial time if it runs in time polynomial in $n$, $m$ and $K$. A computational problem that is weakly NP-hard does not have polynomial time algorithms (unless P=NP), but may have pseudo polynomial time algorithms. 

We now discuss the computational  complexity of tasks related to the current paper.

\begin{proposition}
\label{pro:complete}
    The algorithm proving Lemma~\ref{lem:complete} runs in strongly polynomial time, using only comparison queries.
\end{proposition}

\begin{proof}
    Inspection shows that the pre-processing step in the proof of Lemma~\ref{lem:complete} runs in polynomial time and can implemented using only comparison queries. The same applies to the LMMS algorithm.
\end{proof}

In contrast, computing the RMMS value is not as easy.

\begin{proposition}
    \label{pro:NPhard}
    Computing the RMMS value for additive valuations is weakly NP-hard. Likewise, computing  feasible $n$-partitions is weakly NP-hard.
\end{proposition}

\begin{proof}
    For $n=2$ and additive valuations, the RMMS equals the MMS, which is weakly NP-hard to compute (by an immediate reduction from the weakly NP-hard problem of {\em partition}). Likewise, finding a feasible partition is weakly NP-hard. The case of larger $n$ can be reduced to from the case of two agents, by adding $n - 2$ items of exceptionally high value. 
\end{proof}

We leave many questions open.

\begin{enumerate}
   \item Can the RMMS value be computed in pseudo polynomial time? For valuations beyond additive, the answer may of course depend on the query model. 
   \item For RMMS, can a feasible $n$-partition be computed in pseudo polynomial time? This task is not necessarily more complex that that of computing the RMMS value. For example, for MXS and additive valuations, a feasible $n$-partition can be computed in polynomial time, by using a simple variation of the LMMS algorithm (see~\cite{CaragiannisGRSV23}). However, computing the MXS value is weakly NP-hard~\cite{CaragiannisGRSV23} (even for $n=2)$.  
\item What is the complexity of computing an RMMS-allocation? Note that the algorithm proving Observation~\ref{obs:RSFfeasible} (this is steps~1 and~4 of the algorithm that proves Observation~\ref{obs:EFXRSF}) runs in polynomial time, if the RMMS value and feasible $n$-partitions can be computed in polynomial time. So, in sufficiently strong query models, RMMS allocations can be computed in polynomial time.
\item What is the complexity of computing a partial RMMS-allocation that is also EFX? If the RMMS value and feasible $n$-partitions can be computed in polynomial time, then the algorithm proving Observation~\ref{obs:EFXRSF} runs in weakly polynomial time. Are there query models in which such allocations can be found in polynomial time? 
\end{enumerate}


\subsection*{Acknowledgements}

This research was supported in part by the Israel Science Foundation (grant No. 1122/22). I thank Hannaneh Akrami for providing useful comments on a previous version of this manuscript.

\bibliographystyle{alpha}


\begin{thebibliography}{CKM{\etalchar{+}}19}

\bibitem[ABM18]{ABM18}
Georgios Amanatidis, Georgios Birmpas, and Vangelis Markakis.
\newblock Comparing approximate relaxations of envy-freeness.
\newblock In {\em Proceedings of the Twenty-Seventh International Joint
  Conference on Artificial Intelligence, {IJCAI} 2018}, pages 42--48, 2018.

\bibitem[AG24]{AG24}
Arash Ashuri and Vasilis Gkatzelis.
\newblock Simultaneously satisfying {MXS} and {EFL}.
\newblock {\em CoRR}, abs/2412.00358, 2024.

\bibitem[AMNS17]{amanatidis2017approximation}
Georgios Amanatidis, Evangelos Markakis, Afshin Nikzad, and Amin Saberi.
\newblock Approximation algorithms for computing maximin share allocations.
\newblock {\em ACM Transactions on Algorithms (TALG)}, 13(4):1--28, 2017.

\bibitem[AR25a]{AkramiR25a}
Hannaneh Akrami and Nidhi Rathi.
\newblock Achieving maximin share and {EFX/EF1} guarantees simultaneously.
\newblock In {\em AAAI-25, Sponsored by the Association for the Advancement of
  Artificial Intelligence}, pages 13529--13537. {AAAI} Press, 2025.

\bibitem[AR25b]{AkramiR25}
Hannaneh Akrami and Nidhi Rathi.
\newblock Epistemic {EFX} allocations exist for monotone valuations.
\newblock In {\em AAAI-25}, pages 13520--13528, 2025.

\bibitem[BBKN18]{BarmanBMN18}
Siddharth Barman, Arpita Biswas, Sanath Krishnamurthy, and Yadati Narahari.
\newblock Groupwise maximin fair allocation of indivisible goods.
\newblock In {\em Proceedings of the Thirty-Second {AAAI} Conference on
  Artificial Intelligence, (AAAI-18)}, pages 917--924, 2018.

\bibitem[BF22]{BF22}
Moshe Babaioff and Uriel Feige.
\newblock Fair shares: Feasibility, domination and incentives.
\newblock In {\em {EC} '22: The 23rd {ACM} Conference on Economics and
  Computation, 2022}, page 435. {ACM}, 2022.

\bibitem[BLL{\etalchar{+}}24]{bu2024fair}
Xiaolin Bu, Zihao Li, Shengxin Liu, Jiaxin Song, and Biaoshuai Tao.
\newblock Fair division of indivisible goods with comparison-based queries.
\newblock {\em CoRR}, abs/2407.18133, 2024.

\bibitem[BS24]{BarmanS24}
Siddharth Barman and Mashbat Suzuki.
\newblock Compatibility of fairness and nash welfare under subadditive
  valuations.
\newblock {\em CoRR}, abs/2407.12461, 2024.

\bibitem[Bud11]{Budish11}
Eric Budish.
\newblock The combinatorial assignment problem: Approximate competitive
  equilibrium from equal incomes.
\newblock {\em Journal of Political Economy}, 119(6):1061--1103, 2011.

\bibitem[BUF23]{BUF23}
Gilad Ben-Uziahu and Uriel Feige.
\newblock On fair allocation of indivisible goods to submodular agents.
\newblock {\em CoRR}, abs/2303.12444, 2023.

\bibitem[CCMS25]{cCMS25}
George Christodoulou, Vasilis Christoforidis, Symeon Mastrakoulis, and Alkmini
  Sgouritsa.
\newblock Maximin share guarantees for few agents with subadditive valuations.
\newblock {\em CoRR}, abs/2407.05141, 2025.

\bibitem[CGH19]{CaragiannisGH19}
Ioannis Caragiannis, Nick Gravin, and Xin Huang.
\newblock Envy-freeness up to any item with high nash welfare: The virtue of
  donating items.
\newblock In {\em Proceedings of the 2019 {ACM} Conference on Economics and
  Computation, {EC} 2019, Phoenix, AZ, USA, June 24-28, 2019}, pages 527--545.
  {ACM}, 2019.

\bibitem[CGR{\etalchar{+}}23]{CaragiannisGRSV23}
Ioannis Caragiannis, Jugal Garg, Nidhi Rathi, Eklavya Sharma, and Giovanna
  Varricchio.
\newblock New fairness concepts for allocating indivisible items.
\newblock In {\em Proceedings of the Thirty-Second International Joint
  Conference on Artificial Intelligence, {IJCAI} 2023}, pages 2554--2562, 2023.

\bibitem[CKM{\etalchar{+}}19]{CKMPSW19}
Ioannis Caragiannis, David Kurokawa, Herv{\'e} Moulin, Ariel~D Procaccia,
  Nisarg Shah, and Junxing Wang.
\newblock The unreasonable fairness of maximum {Nash} welfare.
\newblock {\em ACM Transactions on Economics and Computation (TEAC)},
  7(3):1--32, 2019.

\bibitem[CKMS21]{ChaudhuryKMS21}
Bhaskar~Ray Chaudhury, Telikepalli Kavitha, Kurt Mehlhorn, and Alkmini
  Sgouritsa.
\newblock A little charity guarantees almost envy-freeness.
\newblock {\em {SIAM} J. Comput.}, 50(4):1336--1358, 2021.

\bibitem[Fei25]{feige25}
Uriel Feige.
\newblock From multi-allocations to allocations, with subadditive valuations.
\newblock {\em CoRR}, abs/2506.21493, 2025.

\bibitem[FH25]{FH25}
Uriel Feige and Shengyu Huang.
\newblock Concentration and maximin fair allocations for subadditive
  valuations.
\newblock {\em CoRR}, abs/2407.13541, 2025.

\bibitem[FST21]{FST21}
Uriel Feige, Ariel Sapir, and Laliv Tauber.
\newblock A tight negative example for {MMS} fair allocations.
\newblock In {\em International Conference on Web and Internet Economics},
  pages 355--372. Springer, 2021.

\bibitem[GHS{\etalchar{+}}22]{GhodsiHSSY22}
Mohammad Ghodsi, Mohammad~Taghi Hajiaghayi, Masoud Seddighin, Saeed Seddighin,
  and Hadi Yami.
\newblock Fair allocation of indivisible goods: Beyond additive valuations.
\newblock {\em Artif. Intell.}, 303:103633, 2022.

\bibitem[KPW18]{KPW18}
David Kurokawa, Ariel~D. Procaccia, and Junxing Wang.
\newblock Fair enough: Guaranteeing approximate maximin shares.
\newblock {\em J. {ACM}}, 65(2):8:1--8:27, 2018.

\bibitem[LMMS04]{LiptonMMS04}
Richard~J. Lipton, Evangelos Markakis, Elchanan Mossel, and Amin Saberi.
\newblock On approximately fair allocations of indivisible goods.
\newblock In {\em Proceedings of the 5th {ACM} Conference on Electronic
  Commerce}, EC'04, pages 125--131, 2004.

\bibitem[SS25]{SS25}
Masoud Seddighin and Saeed Seddighin.
\newblock Beating the logarithmic barrier for the subadditive maximin share
  problem.
\newblock {\em CoRR}, abs/2506.05613, 2025.

\end{thebibliography}

\newcommand{\etalchar}[1]{$^{#1}$}

\end{document}